\documentstyle[prl,multicol,aps]{revtex}
\begin{document}
\preprint{SNUTP 96-022}
\title{Coulomb Gaps in One-Dimensional Spin-Polarized Electron Systems}
\author{Gun Sang Jeon and M.Y. Choi}
\address{Department of Physics and Center for Theoretical Physics\\
     Seoul National University\\
     Seoul 151-742, Korea}
\author{S.-R. Eric Yang}
\address{Department of Physics\\
         Korea University\\
         Seoul 136-701, Korea}
\maketitle
\draft
\begin{abstract}
We investigate the density of states (DOS) near the Fermi
energy of one-dimensional spin-polarized electron systems in the quantum regime
where the localization length is comparable to or 
larger than the inter-particle distance.
The Wigner lattice gap of such a 
system, in the presence of weak disorder, can occur precisely at the
Fermi energy, coinciding with the Coulomb gap in position. 
The interplay between the two is investigated by
treating the long-range Coulomb interaction and the random 
disorder potential in a self-consistent Hartree-Fock approximation. 
The DOS near the Fermi energy is found to be well described by 
a power law whose exponent decreases with increasing disorder strength.
\end{abstract}
\thispagestyle{empty}
 
\begin{multicols}{2}
The effects of electron interactions are particularly
strong in one-dimensional (1D) systems, leading to Luttinger liquids
or to various instabilities \cite{Sol}.  Random disorder also has
strong effects in 1D systems: Regardless of disorder
strength, all states are known to be localized with the localization
length comparable to the mean free path \cite{L&R}.
The interplay of electron interactions and random
disorder in low-dimensional systems is of great 
current interest \cite{B&K,K&F}.
It has been shown recently
that in the absence of disorder the long-range Coulomb interaction between
electrons reduces quantum fluctuations
so that the ground state acquires quasi-long range order much close to
a 1D Wigner crystal \cite{S}.  This reduction of quantum
fluctuations is reflected in the density of states (DOS), which vanishes
faster than any power as the Fermi energy $\mu$ is approached~\cite{S,F&B}:
\begin{equation} \label{Lgap}
  g(E)\sim \exp\{-A\,[\,\ln(E_c/|E-\mu|)\,]^{3/2}\},
\end{equation}
where $A$ and $E_c$ are appropriate constants.
In the strongly localized regime, where 
the overlap of electron wave functions is negligible,
electrons may be treated classically and the DOS
exhibits a Coulomb gap of the form \cite{Efros}:
\begin{equation} \label{Cgap}
  g(E) \sim [\,\ln(E_c/|E - \mu|)\,]^{-1}.
\end{equation}
When the Thouless length is shorter than the localization length,
the system is in the disordered Fermi liquid regime, where the overlap of
electron wave functions can be significant.
In this case the electron interactions can be included
perturbatively within the weak localization theory~\cite{AA},
and the first-order correction to the DOS near the Fermi
level is given by~\cite{BKK}
\begin{equation}
\delta g(E) \sim |E- \mu|^{-1/2},
\end{equation}
except in the limit 
$E\rightarrow\mu$, where the perturbation theory is 
expected to breakdown.

In this work we employ a self-consistent Hartree-Fock (HF) method 
to investigate the DOS
at zero temperature in the quantum regime where the 
localization length is comparable to or
larger than the inter-particle distance~\cite{Y&M}.
Both the long-range Coulomb 
interaction and disorder are expected to reduce quantum fluctuations:
The Coulomb interaction
pushes the system to the classical limit, where quantum fluctuations
can be neglected except for the redefinition of the strength of
the impurity potential~\cite{M&G}, and
disorder is expected to restore the Fermi liquid behavior~\cite{Naon,Y&A}.
Thus a HF mean-field approximation may provide a reasonable description of 
the interplay between disorder and the Coulomb interaction~\cite{HF}.
This interplay is especially interesting
with spin-polarized electrons since the Fermi wave vector is 
equal to $\pi/a$, where $a$ is the period of
the Wigner lattice.  In the presence of weak disorder we
expect that
the processes leading to the Coulomb gap interact with 
the Bragg scattering leading to a gap in the DOS of a Wigner crystal.
Such a spin-polarized system of electrons can be realized in 
organic chains~\cite{Sol}
or in quasi-one-dimensional quantum wires in a strong magnetic 
field~\cite{Y&A}.  
We find in the quantum regime that the DOS can 
be fitted well with a power-law, and 
that the exponent decreases as the strength of disorder is increased.

A prototype model in which the interplay between
disorder and the Coulomb interaction can be investigated is 
a 1D jellium model 
with $N$ spin-polarized electrons, interacting with 
each other via the long-range Coulomb interaction in the presence of random 
impurities.
We perform a finite-size calculation in a cell of
length $L$ with periodic boundary conditions.  When the effects of
the image charges are taken into account the Coulomb interaction can be
written as
\begin{equation}
V_{C}(x_{1}-x_{2})= \sum_{l=-\infty}^{\infty}\frac{e^{2}}{\epsilon
[(x_{1}-x_{2}-lL)^{2}+d^{2}]^{1/2}},
\end{equation}
where $d$ is the cut-off length corresponding to the transverse
dimension of the system~\cite{S} and $\epsilon$ is the dielectric constant.
For simplicity, we assume that the impurity at
position $x_i$ is characterized by the $\delta$-function
potential with random strength $W_i$.  This gives the total impurity
potential of $N_I$ impurities in the form
\begin{equation}
W_I(x)=\sum_{l=-\infty}^{\infty} \sum_{i=1}^{N_I}  W_i \delta(x-x_i-lL),
\end{equation}
where $W_i$ and $x_i$ are quenched random variables distributed
uniformly in the range $[-W_{max}/2,\, W_{max}/2]$ and in the
interval $[-L/2, \,L/2]$, respectively. 
The electron-electron interaction is treated self-consistently
within a HF approximation in the momentum space.
We expand the HF single-particle wavefunction for state $\alpha$ as follows: 
\begin{equation}
\Psi_{\alpha}(x)=\sum_{k}c_{k\alpha}\phi_{k}(x),
\end{equation} 
where the basis states are
$$
\phi_{k}(x)= \frac{1}{\sqrt{L}}\exp(ikx)
$$
and the wavevector $k$ takes integer multiples of $2\pi/L$.  
The expansion coefficients $c_{k\alpha}$ satisfy the equations
\begin{eqnarray}
\sum_{k'}\langle k |H_{0}+W_{I}+U+V_{HF}|k'\rangle c_{k'\alpha}=
\epsilon_{\alpha} c_{k\alpha},
\label{HF}
\end{eqnarray}
where the matrix elements are evaluated by integrating the coordinate over
the interval $[-L/2, \,L/2]$.  Here $H_{0}$ is the kinetic energy 
and the HF matrix elements are given by
\begin{equation}
\langle k |V_{HF}|k'\rangle=\sum_{\alpha}n_{\alpha}
[\langle k,\alpha |V_{C}|k',\alpha \rangle
-\langle k,\alpha |V_{C}|\alpha,k'\rangle],
\label{HF2}
\end{equation}
where the first and second terms are the Hartree and exchange potentials, 
respectively.  The potential $U$ is due to the uniform positive 
background charge and $n_{\alpha}$ is $1/0$ 
for occupied/unoccupied states $\alpha$.
The HF matrix elements in Eq.~(\ref{HF2}) depend on the expansion
coefficients via
\begin{equation}
\langle k,\alpha |V_{C}|k',\alpha \rangle=
\sum_{k_{1},k_{2}}c_{k_{1}\alpha}^{*}c_{k_{2}\alpha}\langle k,k_{1} |V_{C}|
  k',k_{2}\rangle \nonumber
\end{equation}
and 
\begin{equation}
\langle k,\alpha |V_{C}|\alpha,k' \rangle=
\sum_{k_{1},k_{2}}c_{k_{1}\alpha}^{*}c_{k_{2}\alpha}\langle k,k_{1} |V_{C}|
k_{2},k'\rangle, \nonumber
\end{equation}
where the Coulomb matrix elements are given by
\begin{equation}
\langle k_{1},k_{2} |V_{C}|k'_{1}k'_{2}\rangle=
\delta _{k_{1}+k_{2},k'_{1}+k'_{2}}
K(k_{1}-k'_{1}) \nonumber
\end{equation}
with 
\begin{equation}
K(k)=\frac{1}{L} \sum_{l=-\infty}^{\infty}\int_{-L/2}^{L/2} dx
\frac{e^{2}}{\epsilon[(x-lL)^{2}+d^{2}]^{1/2}}\exp(-ikx). \nonumber
\end{equation}

In our numerical work, the electron number $N$ is fixed to $L/d$, 
so that the inter-particle distance is equal to $d$.
We measure the strength of 
the Coulomb interaction and impurity potential using dimensionless
parameters $V=V_{I}/E_{K}$ and $W=V_{D}/E_{K}$, where
$V_{I}=e^{2}/\epsilon d$, $V_{D}=W_{max}/d$, and 
$E_{K} \equiv \hbar^{2}k_{0}^{2}/2m_{e}$ with $k_{0} \equiv 2\pi/d$.
The impurity number $N_{I}$ is set equal to $5N$ while
the dimension of the Hamiltonian matrix is taken to be 
$2N{+}1$ for $W=0.1$, $3N{+}1$ for $W=0.4$, and $5N{+}1$ for $W=1.0$. 
The convergence of
the obtained results has been tested by increasing 
the dimension of the matrix.
Figure 1 displays the probability distribution function of an electron at the
Fermi energy in strongly localized ($W=1.0$), intermediate ($W=0.1$), 
and disorder free ($W=0$) regimes.
We see that the localization length decreases with the disorder strength, 
and particularly, in the absence of impurities, observe 1D 
crystalline order rather than an electron liquid. 
This mean-field result is in qualitative agreement with the result 
obtained through the use of the bosonization technique, which has predicted
quasi-long-range-order with the correlation function
decaying slower than any power, regardless of
the strength of the long-range Coulomb interaction~\cite{S}.
At $W=0.1$ the electrons are in the quantum regime  
since the localization length extends over several times the inter-particle
distance.  We note that even at $W=1.0$ the electrons 
the localization length is comparable to 
the inter-particle distance. 

The average DOS integrated over the system size is given by
\begin{equation} \label{D}
 \bar{g}(E) \equiv \frac{\langle\!\langle \Delta D(E) 
               \rangle\!\rangle }{\Delta E},
\end{equation}
where $\Delta D(E)$ represents the number of eigenenergies 
in the energy range of width $\Delta E$ around energy
$E$ and the double angular brackets denote the disorder average.
Here we set the value of $\Delta E$ to be $0.05$ in units of
$E_{K}$, and compute the DOS by means of Eq.~(\ref{D}).
Figure 2 displays $\bar{g}(E)$ at $W=0.1$ for five different 
system sizes $L/d=30$, $40$, $50$, $60$, and $80$.
We observe that a large reduction of the DOS occurs 
near the Fermi energy.
Since there is an energy range where no 
significant DOS is present, we need to 
average over many disorder realizations to get accurate 
values of the DOS.  We find that for $W=0.1$ the number $N_D$
of disorder realizations
between 1001 and 3844 is quite sufficient.
To analyze the shape of the gap quantitatively,
we fit the DOS near the Fermi energy to a power law:
\begin{equation}
  g(E) \propto |E-\mu|^{\alpha}, 
\end{equation}
where the DOS $g(E)$ is related to the obtained (average) DOS 
$\bar{g}(E)$ via
\begin{equation}
  \bar{g} (E) = {1\over \Delta E} 
                   \int_{E-\Delta E/2}^{E+\Delta E/2} dE\, g(E).
\end{equation}
This fit of the DOS for $W=0.1$
is shown in Fig.~3.  
The power law fits our numerical results quite well,
with $\mu = -0.29 \pm 0.02$ and $\alpha = 5.80 \pm 0.16$.  
For comparison, we have also fitted $\bar{g}(E)$ 
to a logarithm, given by Eq.~(\ref{Cgap}), only to find substantial
disagreement: The function, Eq.~(\ref{Cgap}), is a sublinear 
function of $E$ while the numerical data form a superlinear function;
see Fig.~2.         

Figure 4 shows the system size dependence of the exponent $\alpha$
for two different values of disorder strength, $W = 0.1$ and $0.4$. 
The values of $(L/d,\,N_{D})$ for $W=0.1$ are as in Fig.~2, while for $W=0.4$
four new values are chosen.  The numbers of disorder realizations
for $W=0.4$, which have been chosen between $199$ and $477$, 
can be much smaller than those for $W=0.1$ since there are many more states 
near the Fermi energy due to stronger disorder.
The obtained results of the size 
dependence tend to bend up in the large
system, and yield
finite extrapolated values of $\alpha$ in the limit $L/d \rightarrow \infty$, 
which provides an evidence for the algebraic gap formation in the
thermodynamic limit.
As expected, the exponent $\alpha$ decreases with $W$, implying that
the gap softens as the strength of disorder is increased.  
On the other hand, we expect that in the very clean limit ($W \ll 0.1$) 
the power law will fail, leading to the gap of the form in Eq.~(\ref{Lgap}). 

In summary, we have investigated the density of states of 
interacting spinless electrons in the presence of 
random disorder in one dimension. 
We find numerical evidence for the formation of a 
gap near the Fermi level:  
The density of states can be fitted with a power law in the 
quantum regime, where the localization length is comparable to or larger
than the inter-particle distance.
The magnitude of the exponent describing the algebraic gap 
decreases with increasing disorder.

G.S.J. and M.Y.C. acknowledge the partial support from the Basic Science 
Research Institute Program, Ministry of Education of Korea and from the 
Korea Science and Engineering Foundation through the SRC Program, and
thank the Central Computing Center for Education for granting the computing
time of the SP2 supercomputer system.
S.R.E.Y. was supported in part by the Non Directed Research Fund, Korea 
Research Foundation, and thanks A.H. MacDonald for useful conversations.

\end{multicols}
 
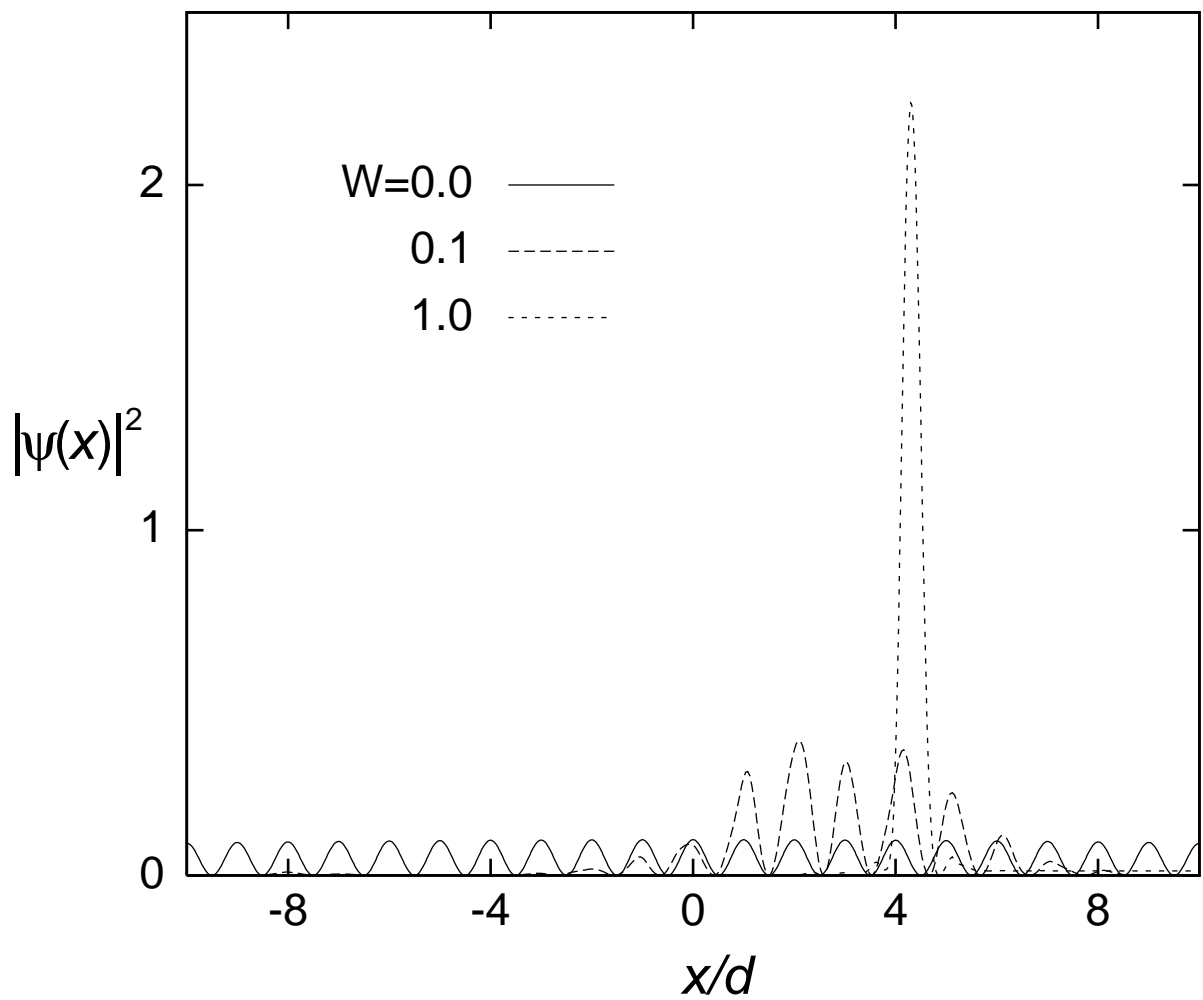
\begin{figure}
\caption{Probability distribution function, in units of $d^{-1}$, 
  of an electron at the Fermi energy for three different 
  values of disorder strength.}
\end{figure}
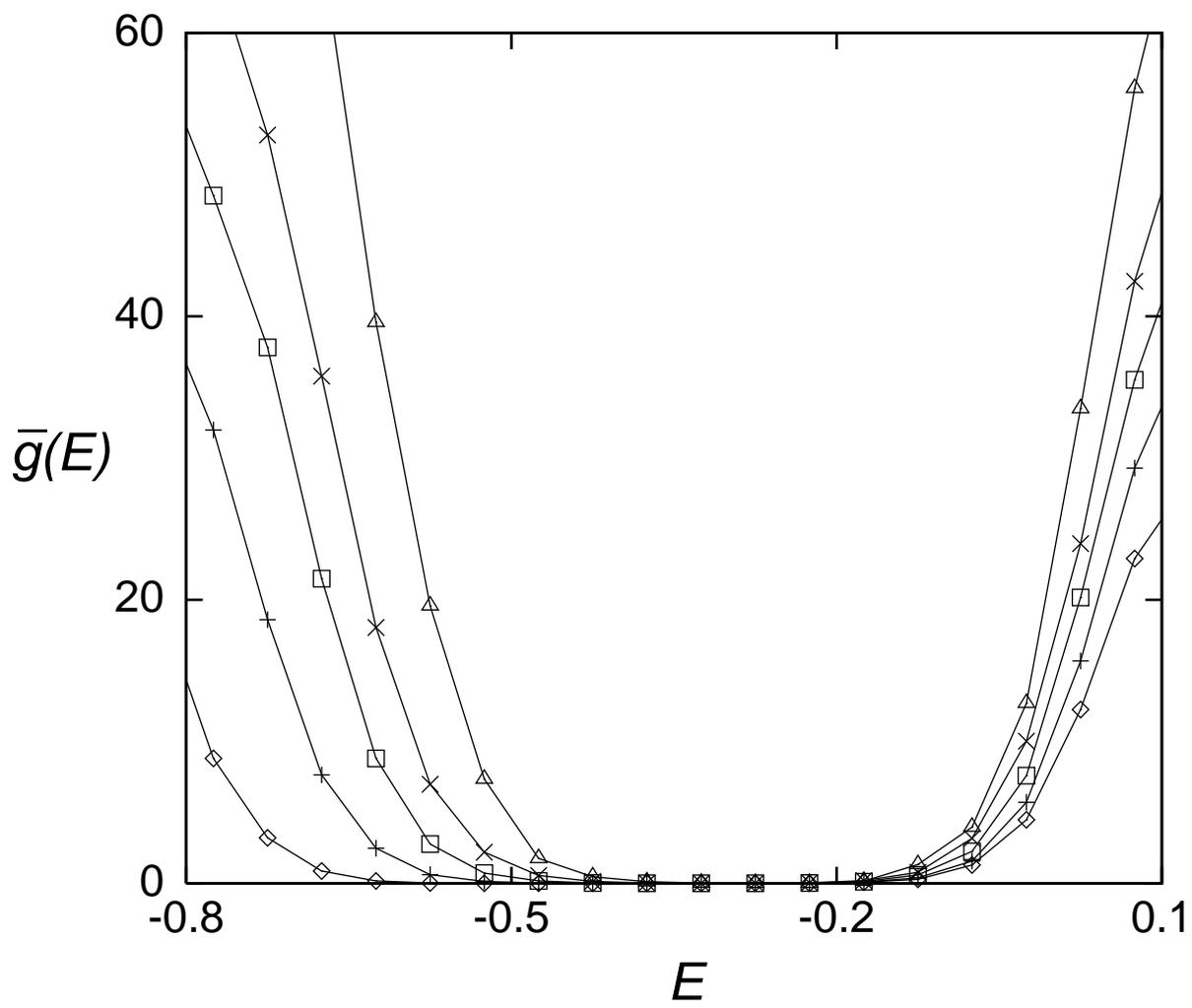
\begin{figure}
\caption{Density of states as a function of energy for various system sizes.  
  The data marked by diamonds($\Diamond$), plus signs($+$), squares($\Box$),
  crosses($\times$), and triangles($\triangle$) correspond to
  $(L/d,\,N_{D})$=(30, 1924), (40, 3740), (50, 3108), (60, 3844) and 
  (80, 1001), respectively.  The disorder strength
  $W$ and the electron density $n$ are fixed to $0.1$ and $1/d$, respectively.}
\end{figure}
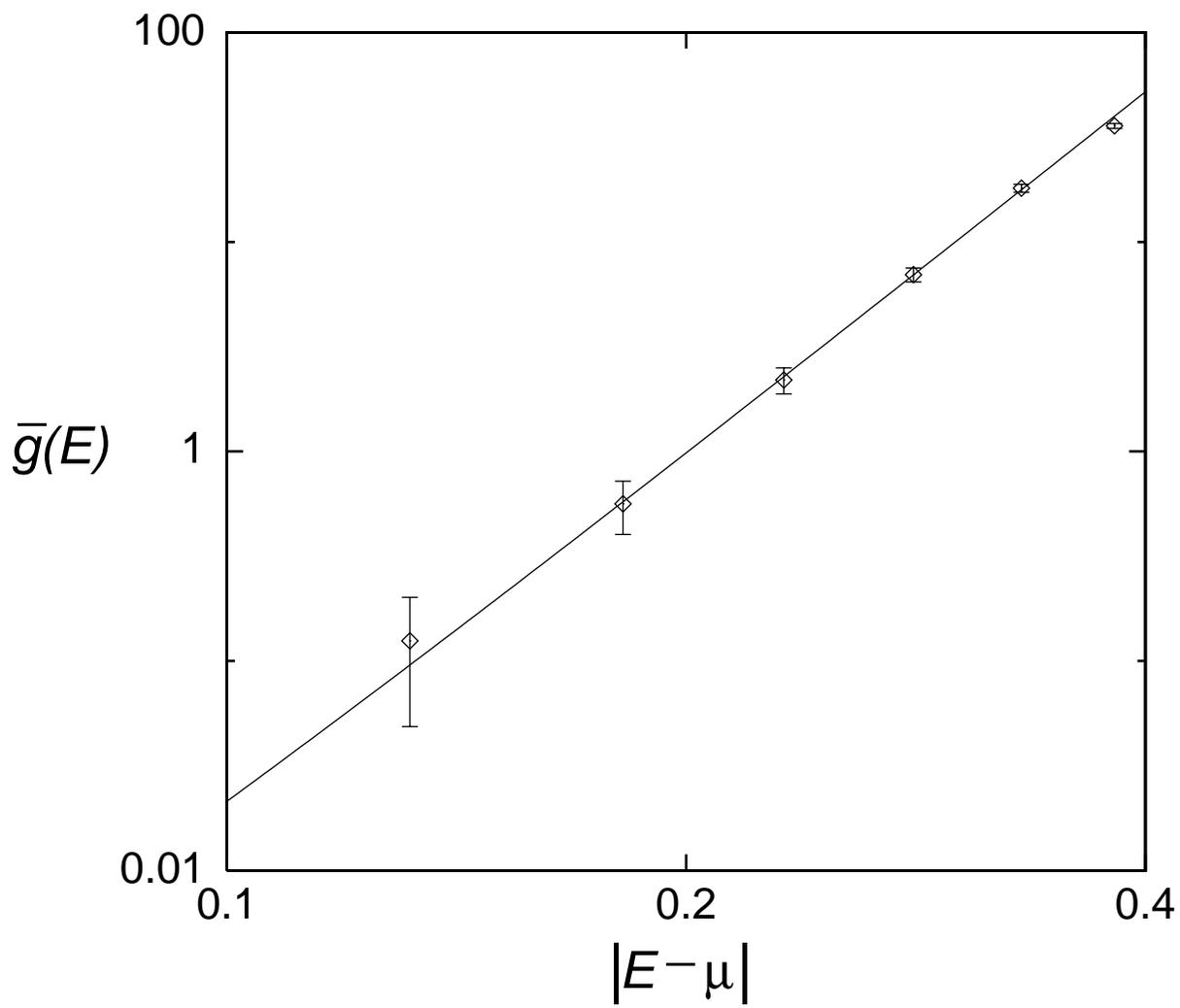
\begin{figure}
\caption{Power-law fit of $\bar{g}(E)$ in a system of size $L/d=60$ and 
  disorder strength $W=0.1$.}
\end{figure}
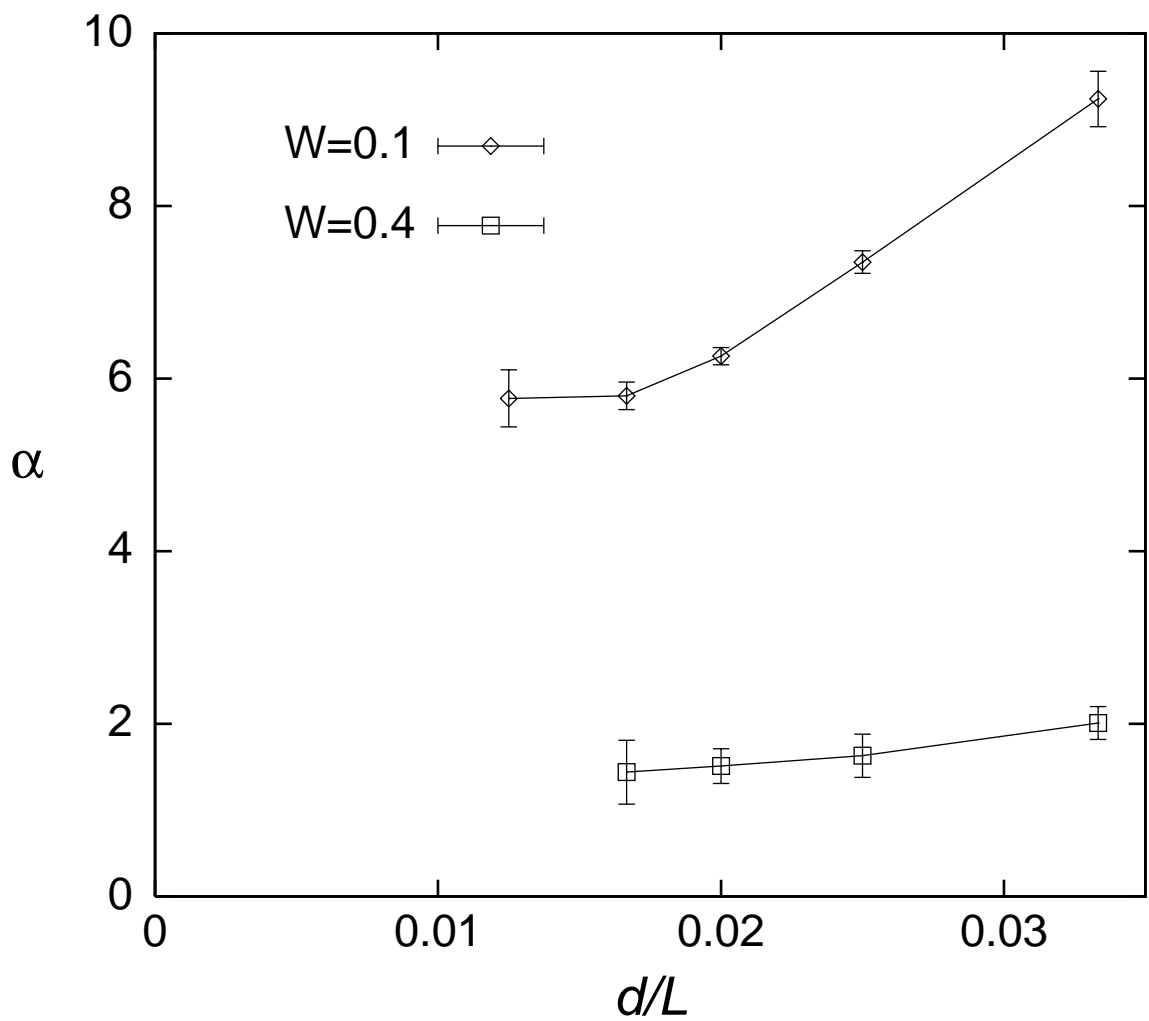
\begin{figure}
\caption{Exponent $\alpha$ as a function of the inverse of the
  system size for two values of disorder strength, $W = 0.1$ and $0.4$.
  While the parameters $(L/d,\,N_{D})$ for $W = 0.1$ are the same as those 
  in Fig.~2, for $W=0.4$ we have used $(L/d,\,N_{D})$=(30, 477), (40, 296), 
  (50, 252), and (60, 199).  The electron density is fixed to $1/d$.
  The lines are guides to the eye.}
\end{figure}


\begin{references}
\bibitem{HF}  The effects beyond the HF approximation may be estimated by 
 screening the exchange interaction within the GW approximation. 
 Since the electrons are all localized here,
 the screened interaction will be essentially identical to the bare 
 interaction except for the simple change in the dielectric constant 
 $\epsilon$.  Thus we expect that the GW approximation would not yield 
 qualitatively different results.
 Indeed the HF calculation is known to be exact in the strong-disorder limit,
 regardless of the dimension.  See Ref.~[10].
\end{references}
\end{document}